\documentclass[11pt]{article}

\usepackage{amsmath,amssymb,setspace}

\usepackage{graphicx}               
\usepackage{url}
\usepackage{xcolor}

\numberwithin{equation}{section}

\renewcommand\appendix{\par
  \setcounter{section}{0}
  \setcounter{subsection}{0}
  \setcounter{figure}{0}
  \setcounter{table}{0}
  \renewcommand\thesection{\Alph{section}}
  \renewcommand\thefigure{\Alph{section}\arabic{figure}}
  \renewcommand\thetable{\Alph{section}\arabic{table}}
}

\def\lsim{\mathrel{\raise.3ex\hbox{$<$\kern-.75em\lower1ex\hbox{$\sim$}}}}
\def\gsim{\mathrel{\raise.3ex\hbox{$>$\kern-.75em\lower1ex\hbox{$\sim$}}}}
\def\half{\tfrac{1}{2}}
\def\lsub#1{_{\lower 1.5pt\hbox{$\scriptstyle#1$}}}

\title{
\vspace*{-2.3cm}
\begin{flushright}
\normalsize{
SCIPP-13/04
}
\end{flushright}
\vspace{1.5cm}
\Large
\textbf{
Decoupling of the Right-handed Neutrino Contribution to the Higgs Mass in Supersymmetric Models
\\
}\vspace*{1.0cm}
}

\author{Patrick Draper and Howard E.~Haber
\vspace{5mm}
\\
\normalsize\emph{Santa Cruz Institute for Particle Physics, University of California, Santa Cruz CA 95064}
}

\begin{document}
\setcounter{page}{0}
\date{\vspace{-5ex}}
\maketitle

\vspace*{1cm}
\begin{abstract}

Recently, it has been argued that in the supersymmetric extension of the seesaw-extended Standard Model,
heavy right-handed neutrinos and sneutrinos may give corrections as large as a few GeV to the mass of the lightest neutral CP-even Higgs boson, even if the soft
supersymmetry-breaking parameters are of order the electroweak scale. The presence of such large corrections would render precise Higgs masses incalculable from measurable low-energy parameters. We show that this is not the case: decoupling is preserved in the appropriate sense and right-handed (s)neutrinos, if they exist, have negligible impact on the physical Higgs masses. \end{abstract}

\thispagestyle{empty}
\newpage

\vspace{-3cm}

\setcounter{footnote}{0} \setcounter{page}{2}
\setcounter{section}{0} \setcounter{subsection}{0}
\setcounter{subsubsection}{0}


\section{Introduction}

The discovery of a new boson near 126 GeV~\cite{:2012gu} that resembles the Higgs boson of the Standard Model has stimulated considerable theoretical interpretation. In supersymmetric models, the observed mass is particularly interesting.  Whereas 126 GeV is compatible with expectations for the mass ($m_h$) of the lightest neutral CP-even Higgs boson of the Minimal Supersymmetric Standard Model (MSSM), large quantum corrections are indicated in order to raise $m_h$ to a value 40\% above $m_Z$~\cite{Heinemeyer:2011aa}.  In the next-to-minimal model (NMSSM), an additional tree-level contribution may also boost the value of $m_h$, but radiative effects are still necessary unless the tri-linear coupling of the singlet and doublet Higgs fields in the superpotential is large~\cite{Hall:2011aa}. Thus the measured Higgs mass provides an important clue to the parameters of the supersymmetric model.

The program of precision calculations of the lightest Higgs boson mass in the MSSM began with one-loop results, given in~\cite{Haber:1990aw}, followed by two-loop contributions given in~\cite{Carena:1995bx}. Partial three-loop results are now available~\cite{Martin:2007pg}. In cases with a large hierarchy between the weak scale and the scale of the stop squarks, resummation has been used to obtain precise results, now at the level of three-loop $\beta$-functions in some cases~\cite{Degrassi:2012ry}.  Residual theoretical uncertainty estimates vary depending on the type of calculation performed, but in the fixed-order case are perhaps of the order 1 GeV for light spectra below a TeV, and 2-3 GeV for heavier spectra~\cite{Martin:2007pg}.

The utility of the computations described above rely on decoupling~--~very heavy states that do not receive their masses from electroweak symmetry breaking are expected to give negligible contribution to the lightest Higgs mass. Only a limited set of model parameters,
 which are in principle accessible at future collider experiments, are thought to be required for an accurate calculation of $m_h$. On the other hand, if an inaccessible heavy sector could provide a significant contribution to $m_h$, then only the size of this contribution could be constrained by comparing the measured $m_h$ to the calculation in terms of observable parameters. This would clearly be a much weaker position.

Recently, it has been suggested in Ref.~\cite{Heinemeyer:2010eg} that in
the seesaw-extended MSSM \cite{Grossman:1997is,mssm-seesaw}, a
right-handed neutrino and sneutrino provides an example of such a
non-decoupling heavy sector, potentially shifting the MSSM prediction
for $m_h$ by as much as a few GeV at one-loop order, even if the soft
supersymmetry (SUSY) breaking parameters remain at the TeV scale. It
was further argued that the large terms appear at order $p^2$ in the
relevant two-point functions, which are invisible to effective
potential estimates that are based on calculations performed at zero
external momenta.\footnote{These contributions to $m_h$ are therefore
quite distinct from corrections that have been found in certain
parameter ranges of the NMSSM with TeV-scale right-handed
neutrinos~\cite{Wang:2013jya}.}

In light of its importance for the interpretation of the observed Higgs boson with $m_h\simeq 126$~GeV, we have performed a reanalysis of the right-handed neutrino and sneutrino contributions to $m_h$ in the seesaw-extended MSSM. We find that the corrections
to $m_h$ due to physics at the seesaw scale are always minuscule, of the order of a billionth of an eV.  This decoupling behavior is manifest in
renormalization schemes in which the $\tan\beta$ counterterm is completely insensitive to phenomena at scales well above the SUSY-breaking scale. One class of decoupling schemes includes physical schemes, where the $\tan\beta$ counterterm is controlled, for example, by the radiative corrections to the mass of the heavy Higgs boson, or by corrections to the decays of the heavy Higgs bosons to down-type fermions. Another class of decoupling schemes subtracts non-decoupling terms by hand, mocking up the behavior of minimal subtraction schemes where heavy particles are fully integrated out at their thresholds and are absent from the low-energy theory.
If a non-decoupling renormalization scheme is employed in the definition of $\tan\beta$, then the decoupling of high-scale physics phenomena in the radiatively-corrected Higgs mass is recovered once $\tan\beta$ is directly related to a low-energy observable.
That is, $\tan\beta$ should be regarded as an intermediary quantity,
which one is free to define in any scheme.  Independently of how one
defines $\tan\beta$,
the MSSM Higgs mass ultimately depends solely on parameters that can
be fixed by experimental measurements at energy scales of order the
SUSY-breaking scale and below. Contributions to the Higgs mass from
energy scales significantly above the SUSY-breaking scale must be negligible.

This paper is organized as follows. In Section 2 we review the
computation of the physical masses of the neutral CP-even Higgs bosons
in the MSSM at one-loop order. We provide compact formulae and discuss
the role of $\tan\beta$ renormalization in the results. In Section 3
we calculate the leading contributions to the lightest Higgs boson
mass from the left and right handed neutrino/sneutrino sectors. We
reduce the full diagrammatic result of Ref.~\cite{Heinemeyer:2010eg}
to simple, approximate analytic formulae in two different
renormalization schemes, and find that in both cases the right-handed
neutrino sector exhibits appropriate decoupling. We provide an
interpretation of our approximate formulae in the more natural setting
of effective field theory. Finally, we study the full one-loop results
numerically, finding again that contributions from the right-handed
neutrino sector are negligible.  Our conclusions are presented in
Section 4.  Explicit expressions for self-energy functions, tadpoles
and the $\tan\beta$ counterterm, which can provide potential
non-decoupling contributions in the computation of the Higgs mass, are
exhibited in Appendix A. Using these approximate forms, one can check
that the non-decoupling terms cancel exactly in the expressions for
the one-loop radiatively-corrected Higgs mass when a suitable
definition of the $\tan\beta$ counterterm is employed.

\section{Physical Higgs Masses at One Loop in the MSSM}

We begin our discussion with a review of the one-loop physical Higgs masses in the MSSM with the minimally required two-Higgs doublet Higgs sector. The neutral field content is
\begin{align}
H_{u,d}^0\equiv\frac{\phi_{u,d}^r+i\phi_{u,d}^i}{\sqrt{2}}+v_{u,d}\,
\end{align}
where $v^2\equiv v_u^2+v_d^2=(174~{\rm GeV})^2$.  Here $v_u$ [$v_d$] are the vacuum expectation values (vevs) of the neutral Higgs fields that couple exclusively to the up-type [down-type]
quark and lepton fields.

The MSSM Higgs scalar potential is given by
\begin{align}
V=m_1^2H_d^\dagger H_d+m_2^2H_u^\dagger H_u-b(H_d H_u+{\rm h.c.})+\tfrac{1}{8}G^2(H_u^\dagger H_u-H_d^\dagger H_d)^2+\tfrac{1}{2}g^2|H_d^\dagger H_u|^2\; ,
\end{align}
where $G^2\equiv g_1^2+g_2^2$, $m_1^2\equiv m_{d}^2+|\mu|^2$, $m_2^2\equiv m_{u}^2+|\mu|^2$, $\mu$ is the supersymmetric Higgs mass parameter, and $m_u^2$, $m_d^2$, and $b$ are soft SUSY-breaking squared-mass parameters.
The linear terms in the potential are given by:
\begin{align}
T_u&\equiv\frac{\partial V}{\partial \phi_u^r}\bigg|_{\phi=0}=\frac{v_u}{\sqrt{2}}\left(2m_2^2+\tfrac{1}{2}G^2(v_u^2-v_d^2)-2b\frac{v_d}{v_u}\right)\;,
\label{baretad1}
\end{align}
\begin{align}
T_d&\equiv\frac{\partial V}{\partial \phi_d^r}\bigg|_{\phi=0}=\frac{v_d}{\sqrt{2}}\left(2m_1^2+\tfrac{1}{2}G^2(v_d^2-v_u^2)-2b\frac{v_u}{v_d}\right)\;.
\label{baretad2}
\end{align}
The quadratic terms yield $2\times2$ scalar and pseudoscalar squared-mass matrices [in the $(\phi_d,\phi_u)$ basis],
\begin{align}
\frac{\partial^2V}{\partial\phi_a^r\partial\phi_b^r}\equiv\mathcal{M}^2_e=\left(
          \begin{array}{cc}
            m_1^2+\tfrac{1}{4}G^2(3v_d^2-v_u^2) & -\tfrac{1}{2}G^2 v_uv_d-b \\
            -\tfrac{1}{2}G^2 v_uv_d-b & m_2^2+\tfrac{1}{4}G^2(3v_u^2-v_d^2) \\
          \end{array}
        \right)\;,
\end{align}

\begin{align}
\frac{\partial^2V}{\partial\phi_a^i\partial\phi_b^i}\equiv\mathcal{M}^2_o=\left(
          \begin{array}{cc}
            m_1^2+\tfrac{1}{4}G^2(v_d^2-v_u^2) & b \\
            b & m_2^2+\tfrac{1}{4}G^2(v_u^2-v_d^2) \\
          \end{array}
        \right).
\end{align}
All parameters appearing in the above formulae should be interpreted as bare parameters.

It is convenient to require that $v_{u,d}$ are stationary points of the full one-loop effective potential, which
is achieved via the tadpole cancellation conditions,
\begin{align}
T_{u,d}+A_{u,d}=0\;.
\label{eq:tadzero}
\end{align}
The functions $A_{u,d}$ are the one-loop tadpole diagrams at zero external momentum, and the $T_{u,d}$ are functions of the bare parameters given in Eqs.~(\ref{baretad1}) and (\ref{baretad2}). Using Eq.~(\ref{eq:tadzero}), the pseudoscalar mass matrix simplifies to
\begin{align}
\mathcal{M}^2_o=\left(
          \begin{array}{cc}
            b\displaystyle\frac{v_u}{v_d}-\displaystyle\frac{A_d}{\sqrt{2}v_d} & b \\[10pt]
            b &        b\displaystyle\frac{v_d}{v_u}-\displaystyle\frac{A_u}{\sqrt{2}v_u} \\
          \end{array}
        \right).
\end{align}
Diagonalizing this matrix and expanding to leading order in $A_{u,d}$, the bare masses for the pseudoscalar~$A$ and the Goldstone boson $G$ are found:
\begin{align}
\label{eq:mabare}
m_{A}^2&=\frac{v^2}{v_uv_d}b-\frac{v_u^2}{v^2}\frac{A_d}{\sqrt{2}v_d}-\frac{v_d^2}{v^2}\frac{A_u}{\sqrt{2}v_u}\;,\\
m_{G}^2&=-\frac{1}{\sqrt{2}v^2}\left(A_dv_d+A_uv_u\right)\;.
\end{align}
Solving Eqs.~(\ref{eq:tadzero}) and (\ref{eq:mabare}) for $b$, $m_1^2$ and $m_2^2$ yields
\begin{align}
b&=\left(\frac{v_uv_d}{v^2}\right)m_A^2+\left(\frac{v_u}{v}\right)^4\frac{A_d}{\sqrt{2}v_u}+\left(\frac{v_d}{v}\right)^4\frac{A_u}{\sqrt{2}v_d}\;,\nonumber\\
m_1^2&=\left(\frac{v_u}{v}\right)^2m_A^2+\left[\left(\frac{v_u}{v}\right)^4-1\right]\frac{A_d}{\sqrt{2}v_d}+\left(\frac{v_dv_u}{v^2}\right)^2\frac{A_u}{\sqrt{2}v_u}+  \frac{G^2}{4}(v_u^2-v_d^2)\;,\nonumber\\
m_2^2&=\left(\frac{v_d}{v}\right)^2m_A^2+\left(\frac{v_uv_d}{v^2}\right)^2\frac{A_d}{\sqrt{2}v_d}+\left[\left(\frac{v_d}{v}\right)^4-1\right]\frac{A_u}{\sqrt{2}v_u}-  \frac{G^2}{4}(v_u^2-v_d^2)  \;.
\end{align}
Inserting these results into $\mathcal{M}^2_e$, we obtain
\begin{align}
\mathcal{M}^2_e=\left(
          \begin{array}{cc}
          m_A^2s^2_\beta+m_Z^2c^2_\beta+\displaystyle\frac{A_d}{\sqrt{2}v_d}(s^4_\beta-1)+\displaystyle\frac{A_u}{\sqrt{2}v_u}s^2_\beta c^2_\beta & -(m_A^2+m_Z^2)s_\beta c_\beta - \displaystyle\frac{A_u}{\sqrt{2}v_u}c^3_\beta s_\beta -\displaystyle\frac{A_d}{\sqrt{2}v_d}s^3_\beta c_\beta \\[10pt]
          -(m_A^2+m_Z^2)s_\beta c_\beta - \displaystyle\frac{A_u}{\sqrt{2}v_u}c^3_\beta s_\beta -\displaystyle\frac{A_d}{\sqrt{2}v_d}s^3_\beta c_\beta & m_A^2c^2_\beta+m_Z^2s^2_\beta+\displaystyle\frac{A_d}{\sqrt{2}v_d }s^2_\beta c^2_\beta+\displaystyle\frac{A_u}{\sqrt{2}v_u}(c^4_\beta-1)
          \end{array}
        \right)\;,
\end{align}
where $m_Z^2\equiv \tfrac{1}{2}G^2v^2$, $s_\beta\equiv\sin\beta$, and $c_\beta\equiv\cos\beta$.
The squared-mass matrix $\mathcal{M}^2_e$ can be diagonalized to obtain the bare masses $m_{h,H}^2$ for the light
neutral CP-even Higgs boson $h$ and the heavy neutral CP-even Higgs boson $H$.

At this stage, it is convenient to replace the bare masses by physical masses:
\begin{align}
m_{h,Z,A,H}^2= m_{hP,ZP,AP,HP}^2-\Sigma_{hh,ZZ,AA,HH}(m_{hP,ZP,AP,HP}^2)\;,
\label{eq:pmass}
\end{align}
where the subscript $P$ indicates the corresponding physical parameter.
The $\Sigma$ functions are the real parts of the corresponding
self-energy functions\footnote{
In our notation, the sum of all one-loop Feynman graphs contributing
to the $\phi\phi$ ($\phi=h,A,H$) and $ZZ$ self-energy functions are
denoted by $-iC_{\phi\phi}(p^2)$ and
$iA_{ZZ}(p^2)g_{\mu\nu}+iB_{ZZ}(p^2)p_{\mu\nu}$, respectively, where $p$ is the
four-momentum of the incoming boson. Only $\Sigma_{ZZ}(p^2)\equiv{\rm
  Re~} A_{ZZ}(p^2)$ and $\Sigma_{\phi\phi}(p^2)\equiv{\rm Re~}
C_{\phi\phi}(p^2)$ are needed to define the physical on-shell boson
masses. Note that the opposite sign
choice in the definition of $\Sigma(p^2)$
is sometimes employed in the literature.}
through which parameters from other sectors
of the theory affect the Higgs masses.
At one-loop order, the arguments of $\Sigma_{hh,HH}$ can be
consistently replaced with the corresponding tree-level expressions
for the physical masses,
\begin{align}
m_{ht,Ht}^2=\frac{1}{2}\left(m_Z^2+m_A^2\mp\sqrt{(m_A^2-m_Z^2)^2+4m_A^2m_Z^2\sin^22\beta}\right).
\label{mhtree}
\end{align}

The replacements of Eq.~(\ref{eq:pmass}) largely sidestep the need to
introduce renormalized mass parameters and counterterms in the
calculation of $m_h$ and $m_H$. The only 
explicit counterterms required are associated with the parameters $v_u$ and $v_d$, 
which are divergent because they are fixed to the
vevs of the bare fields $H_{u,d}$. Rescaling the fields
by wave function renormalizations
renders the vevs finite,
\begin{equation}
v_u\rightarrow \mathcal{Z}_{H_u}^{-1/2} v_u = v_u(1-\tfrac{1}{2}\delta\mathcal{Z}_{H_u})\;,\;\;\;\;\;\;\;v_d\rightarrow \mathcal{Z}_{H_d}^{-1/2} v_d = v_d(1-\tfrac{1}{2}\delta\mathcal{Z}_{H_d})\;.
\label{eq:vevrenorm}
\end{equation}
At one-loop order the renormalization of the vevs affects the Higgs masses only through the parameter $\tan\beta\equiv v_u/v_d$, 
which can be replaced by a renormalized parameter and a counterterm that is fixed by Eq.~(\ref{eq:vevrenorm}):
\begin{align}
&\phantom{line} \nonumber \\[-40pt]
\tan\beta & \rightarrow \tan\beta-\delta\tan\beta\;,\label{eq:tbr} \\[-40pt]
& \phantom{line}\nonumber 
\end{align}
where
\begin{align}
&\phantom{line} \nonumber \\[-40pt]
\delta\tan\beta & \equiv\tfrac{1}{2}(\delta\mathcal{Z}_{H_d}-\delta\mathcal{Z}_{H_u})\tan\beta.
\label{eq:dtbZ} \\[-30pt]
& \phantom{line}\nonumber 
\end{align}

Making the substitutions of Eqs.~(\ref{eq:pmass}) and~(\ref{eq:tbr})
and expanding to leading order in the one-loop functions,
we obtain
\begin{align} \label{mhfull}
m_h^2&=m_{ht}^2-\sin(\beta-\alpha)\frac{\sqrt{2}A_h}{v}-\sin^2(\beta+\alpha)\Sigma_{ZZ}(m_Z^2)+\Sigma_{hh}(m_{ht}^2)\nonumber\\
 &\qquad-\cos^2(\beta-\alpha)\Sigma_{AA}(m_A^2)+\sin^2(\beta-\alpha)\Sigma_{GG}(0)-2m_Z^2\cos^2\beta\sin(2(\beta+\alpha))\delta\tan\beta \; ,
\end{align}
and
\begin{align} \label{mhfull2}
m_H^2&=m_{Ht}^2-\cos(\beta-\alpha)\frac{\sqrt{2}A_H}{v}-\cos^2(\beta+\alpha)\Sigma_{ZZ}(m_Z^2)+\Sigma_{HH}(m_{Ht}^2)\nonumber\\
 &\qquad-\sin^2(\beta-\alpha)\Sigma_{AA}(m_A^2)+\cos^2(\beta-\alpha)\Sigma_{GG}(0)+2m_Z^2\cos^2\beta\sin(2(\beta+\alpha))\delta\tan\beta \;,
\end{align}
where $m_{ht,Ht}$ are the tree-level masses of the neutral CP-even
Higgs bosons [cf.~Eq.~(\ref{mhtree})], $m_A$ and $m_Z$ are the
physical masses\footnote{To simplify the typography, we
remove all $P$ subscripts.  However, all masses in the subsequent formulae should now be interpreted as (finite) physical masses.}
(i.e., input parameters taken from experimental measurements),
the angle $\alpha$ is the tree-level mixing angle obtained in the diagonalization of $\mathcal{M}_e^2$, and
$$
A_h\equiv A_u\cos\alpha-A_d\sin\alpha\,,\qquad\quad A_H\equiv A_u\sin\alpha+A_d\cos\alpha\,,
$$
are the tadpoles with respect to the neutral CP-even Higgs mass basis.

In obtaining the formulae in Eqs.~(\ref{mhfull}) and (\ref{mhfull2})
we have used the tree level relation that relates $\alpha$ to the
parameters $\beta$ and $m_A$ (cf.~Eq.~(A.20) of Ref.~\cite{Gunion:1986nh}),
\begin{align} \label{AZ}
m_A^2=-m_Z^2\frac{\sin\bigl(2(\beta+\alpha)\bigr)}{\sin\bigl(2(\beta-\alpha)\bigr)},
\end{align}
as well as the relation between the tadpoles and the Goldstone self-energy imposed by the requirement that the one-loop Goldstone boson mass vanishes,
\begin{align}
\sqrt{2}\,v\Sigma_{GG}(0)=\cos(\beta-\alpha)A_H+\sin(\beta-\alpha)A_h\;.
\end{align}
As a check of our calculation, we note that in the limit of $\beta=\half\pi$ and $m_A>m_Z$ we have $m_{ht}=m_Z$, $m_{Ht}=m_A$, and $\sin(\beta-\alpha)=1$ at tree-level.
In this case, Eqs.~(\ref{mhfull}) and (\ref{mhfull2}) reduce to
\begin{align}
\label{eq:largetb}
m_h^2&= m_Z^2+\Sigma_{hh}(m_{Z}^2)-\Sigma_{ZZ}(m_Z^2)-\frac{A_h}{\sqrt{2}\,v}\;,\nonumber\\
m_H^2&=m_A^2+\Sigma_{HH}(m_{A}^2)-\Sigma_{AA}(m_A^2)\;,
\end{align}
which reproduces the result for $m_h$ obtained in Ref.~\cite{Haber:1990aw}.

From Eqs.~(\ref{mhfull}) and ~(\ref{mhfull2}), we see that the only counterterm appearing explicitly in the Higgs masses is $\delta\tan\beta$. If only a prediction for $m_{h}$ is desired, then $\delta\tan\beta$ can be eliminated in favor of $m_{H}$, and all instances of the renormalized $\tan\beta$ parameter appearing in the self-energies may be consistently replaced at one-loop order by solving the tree level formula Eq.~(\ref{mhtree}) for $\tan\beta$ as a function of $m_H$.  The end result,
\begin{align}
m_h^2=m_A^2+m_Z^2-m_H^2+\Sigma_{hh}(m_{ht}^2)+\Sigma_{HH}(m_{Ht}^2)-\Sigma_{ZZ}(m_Z^2)-\Sigma_{AA}(m_A^2)-\Sigma_{GG}(0)\;,
\label{eq:mhsumrule}
\end{align}
coincides with a sum rule derived first in Ref.~\cite{Berger:1989hg}.

 In the MSSM, the prediction for $m_h$ and $m_H$
  depends on $\tan\beta$ and other MSSM mass parameters (such as $m_A$
  and the top squark mass and mixing parameters).  In particular,
  since $\tan\beta$ appears in the expressions for $m_{ht}$ and
  $m_{Ht}$ [cf.~Eq.~(\ref{mhtree})],
one must define $\delta\tan\beta$ by specifying a subtraction scheme.  In principle any scheme to define the parameter $\tan\beta$ is allowed. In practice, it is preferable to employ a scheme that satisfies decoupling,
in which case $\tan\beta$ can be determined solely from physical measurements that can be carried out in collider experiments.
In contrast, if a non-decoupling scheme is used, then the definition of $\tan\beta$ depends on unknown contributions from inaccessible heavy sectors, in which case the value of $\tan\beta$ (which is needed to
predict $m_h$ and $m_H$) cannot be determined from low-energy experimental measurements.

Of course, in the context of a specific model of high scale physics, one can employ a non-decoupling scheme to define $\tan\beta$ and
then compute the relation of $\tan\beta$ so defined to some specific low-energy observable.  In this case, one can formally eliminate $\tan\beta$ and re-express the MSSM prediction
for $m_h$ and $m_H$ in terms of the corresponding low-energy observable.  This would then provide a prediction for $m_h$ and $m_H$ in terms of parameters that can be determined solely from low-energy measurements.  Following such a procedure,
one finds that the predicted values for $m_h$ and $m_H$ are completely
insensitive to high-scale physics, as expected from the decoupling
properties of quantum field theory (e.g., see Ref.~\cite{Dobado:1997up}).
By employing a definition of $\tan\beta$ that respects decoupling, the insensitivity of the predicted values for $m_h$ and $m_H$ to high scale physics is manifest.

Suppose that there are no schemes in which $\tan\beta$ can be determined from a low-scale measurement. As a simple example, consider the case of high-scale SUSY in the decoupling limit, where all the superpartner masses and $m_A$ are taken very large,
of order $m_{\rm SUSY}\gg m_Z$. In this case decoupling schemes for $\tan\beta$ are not particularly favored over non-decoupling schemes. On the other hand, the observed Higgs mass is no longer a testable prediction, but rather a scheme-dependent constraint on the two unmeasurable parameters $m_{\rm SUSY}$ and $\tan\beta$. Scheme-dependence is not very important in this case as it can simply be absorbed in an unobservable shift of $\tan\beta$. Furthermore, it does not affect the upper bound on $m_h$ for fixed $m_{\rm SUSY}$, which is obtained in the large $\tan\beta$ limit where scheme-dependent terms vanish. For the rest of this work, we will focus on the case in which the MSSM Higgs mass prediction is testable at colliders.

The standard $\overline{\rm DR}$ scheme~\cite{DR}  will not
automatically yield decoupling. However, it can be modified slightly
(m$\overline{\rm DR}$, in the notation of Ref.~\cite{Heinemeyer:2010eg}) to
remove large logarithms by hand. This subtraction reproduces the
result one would obtain at leading-log order with effective field
theory, in which heavy sectors are integrated out by hand at their
thresholds.  Hence, at leading-log order the
  m$\overline{\rm DR}$ scheme respects decoupling. However, beyond
  leading-log, one should also remove non-decoupling non-logarithmic finite terms that are still present in the m$\overline{\rm DR}$ scheme. This can be achieved in an extension of the m$\overline{\rm DR}$ scheme in which \textit{all} contributions from the heavy sector are subtracted.

A scheme that possesses similar properties, denoted by ``DEC'' (for decoupling) below, fixes the Higgs wave function counterterms as follows,\footnote{The choice of evaluating
the $p^2$--derivatives of the self-energies at $p^2=0$ is one of many possible choices.  Employing a different value of $p^2$ would simply yield a $\tan\beta$ definition
that differs at the one-loop level.  In the approximations used in this paper, the difference in the two definitions of $\tan\beta$ is subdominant and can thus be neglected.}
\begin{align}
\left(\delta\mathcal{Z}_{H_d}\right)_{\rm DEC}&=\frac{d\Sigma_{HH}(p^2)}{dp^2}\bigg |_{\alpha=0\,,\,p^2=0}\;,\nonumber\\[10pt]
\left(\delta\mathcal{Z}_{H_u}\right)_{\rm DEC}&=\frac{d\Sigma_{hh}(p^2)}{dp^2}\bigg |_{\alpha=0\,,\,p^2=0}\;.\label{deltaZs}
\end{align}
In this scheme, the $\tan\beta$ counterterm is given by Eq.~(\ref{eq:dtbZ}),
\begin{equation} \label{dtanbdec}
(\delta\tan\beta)_{\rm DEC}=\tfrac{1}{2}\tan\beta\left\{\frac{d[\Sigma_{HH}(p^2)-\Sigma_{hh}(p^2)]}{dp^2}\right\}_{\alpha=0\,,\,p^2=0}\,.
\end{equation}
Indeed, the DEC scheme manifestly removes large logarithms and finite terms from heavy sectors (as we exhibit explicitly in Section 3.1). This subtraction scheme also removes additional contributions that depend on the low-energy sectors
(without affecting the decoupling behavior of the scheme).  In fact,
this is reminiscent of the on-shell scheme (the definition of which
does not involve the limit $\alpha\rightarrow 0$) which was observed
in Ref.~\cite{Heinemeyer:2010eg} to respect decoupling, but was discarded
in favor of the m$\overline{\rm DR}$ scheme, as the latter was deemed
to be more numerically stable.
We emphasize that even with a scheme (such as the DEC scheme) that is not directly related to any particular physical measurement, decoupling is preserved if the effects of the heavy sector that do not vanish in the large mass limit are fully removed by hand. In particular this is how effective field theory analysis should be performed in mass-independent schemes~\cite{Weinberg:1980wa}.

\clearpage

Another possibility is to demand that some physical (measurable) quantity is given
at one-loop order by its tree-level formula. Two such quantities are
the mass $m_H$ and the decay rate $\Gamma(A\rightarrow\tau\tau)$. In
the former case [denoted as the ``HiggsMass" (HM) scheme], the
$\tan\beta$ counterterm is obtained by setting $m_H^2=m_{Ht}^2$
in Eq.~(\ref{mhfull2}), which \textit{defines} $\tan\beta$ in terms of the \textit{low-energy} physical parameters $m_Z$, $m_H$ and $m_A$, so that all one-loop pieces cancel:
\begin{align}
(\delta\tan\beta)_{\rm HM}=\frac{1}{2m_Z^2\cos^2\beta\sin\bigl(2(\beta+\alpha)\bigr)}\bigg(&\cos(\beta-\alpha)\frac{\sqrt{2}A_H}{v}+\cos^2(\beta+\alpha)\Sigma_{ZZ}(m_Z^2)-\Sigma_{HH}(m_{Ht}^2)\nonumber\\
 &+\sin^2(\beta-\alpha)\Sigma_{AA}(m_A^2)-\cos^2(\beta-\alpha)\Sigma_{GG}(0)\bigg)\;.\label{deltatbHM}
\end{align}

A detailed and complementary discussion of $\tan\beta$ renormalization
appears in Ref.~\cite{Freitas:2002um}. In this reference, the authors do not emphasize decoupling properties, but exhibit other flaws among all available schemes. For example, $\overline{\rm DR}$ is gauge-dependent at one-loop, the HM scheme can lead to large perturbative corrections and numerical instability, and using $\Gamma(A\rightarrow\tau\tau)$ is both technically complicated and introduces flavor dependence into $\tan\beta$.
For our purposes of exhibiting decoupling in the next section, we will use the DEC and HM schemes as examples.

Regardless of the scheme used to define $\delta\tan\beta$, measuring $\Gamma(A\rightarrow\tau\tau)$ is a good way to experimentally determine the numerical value of renormalized parameter $\tan\beta$ in the given scheme. Once $\tan\beta$, $m_A$, and the soft parameters are fixed (either by hand or from experimental determinations), $m_h$ and $m_H$ become predictions of the theory.

\section{Right-Handed Sneutrino Contributions to $\boldsymbol{m_h}$}
Right-handed neutrinos and sneutrinos obtain supersymmetric masses and couple to the Higgs sector through the following superpotential interactions~\cite{Grossman:1997is,mssm-seesaw}:
\begin{align} \label{superpot}
W=\mu H_dH_u + y_\nu LH_uN-y_lLH_dR+\frac{1}{2}m_MNN\;,
\end{align}
where $N$ and $R$ represent the right-handed neutrino and lepton multiplets, respectively, and $m_M$ is the Majorana mass. There are also new soft SUSY-breaking couplings and masses given by the potential
\begin{align}
V_{\rm soft}=m_{\tilde{R}}^2\tilde{N}^*\tilde{N}+(y_\nu A_\nu H^0_U\tilde{\nu}_L\tilde{N}^*+m_M B_\nu\tilde{N}\tilde{N}+\rm{h.c.})\;.
\end{align}
In general all masses and couplings are $3\times 3$ matrices in flavor space, but for simplicity we consider only a single flavor. The resulting neutrino mass matrix is given by
\begin{align}
\label{eq:mnu}
\mathcal{M}_\nu=
\left(
\begin{array}{cc}
 0 & m_D    \\
 m_D  & m_M
\end{array}
\right)\;,
\end{align}
where $m_D\equiv y_\nu v_u$. The CP-even/odd $(+/-)$ sneutrino mass matrices are given by~\cite{Grossman:1997is}:
\begin{align}
\mathcal{M}^2_{\tilde{\nu}{\pm}}=
\left(
\begin{array}{cc}
 m^2_{\tilde{L}} +m_D^2+\frac{1}{2}m_Z^2\cos 2\beta & \quad m_D (A_\nu-\mu\cot\beta\pm m_M)   \\
 m_D (A_\nu-\mu\cot\beta\pm m_M)  &\quad  m^2_{\tilde{R}} +m_D^2+m_M^2\pm 2B_\nu m_M
\end{array}
\right)\;,
\end{align}
where $m^2_{\tilde{L}}$ is the usual soft-breaking mass for the left-handed sneutrinos present in the MSSM.

In the analysis presented in this paper, we
consider only a single flavor of right-handed
neutrinos and sneutrinos as described above.
Nevertheless, our conclusions should not be
affected by the presence of additional generations of right-handed
neutrinos and sneutrinos.

\subsection{Approximate Diagrammatic Result}

We expect that the right-handed neutrino and sneutrino contributions
to the physical Higgs masses should decouple as the Majorana mass
scale becomes much larger than the soft supersymmetry breaking scales,
if all other parameters are held fixed.  This expectation is based on
the fact that the Majorana mass term $m_M$ that appears in
the superpotential [cf.~eq.~(\ref{superpot})] is a
supersymmetry-preserving parameter.  Indeed it is well known that the
corrections to the tree-level Higgs mass relations in the MSSM are due
entirely to SUSY-breaking effects.  In contrast, we do not expect
decoupling if the SUSY-breaking parameters associated with the
right-handed sneutrino sector are taken very large.  In the
calculations presented in this section, we shall initially assume that all
SUSY-breaking masses are no larger than $\mathcal{O}(1~{\rm TeV})$.
The consequences of large SUSY-breaking in the right-handed sector
will be briefly considered in Section 3.4.

The relevant one-loop tadpoles and self-energy functions are given in
the appendix of Ref.~\cite{Heinemeyer:2010eg}. We have independently
computed $\Sigma_{hh,ZZ}$ and $A_{h}$ in the $\cot\beta\rightarrow0$
limit and found agreement except for the minus signs in front of the
$m_Z^2$ terms in the last and third-to-last lines of Eq.~(81) of
Ref.~\cite{Heinemeyer:2010eg}. Inserting the formulae for the one-loop
tadpoles and self-energy functions into Eqs.~(\ref{mhfull}) and
(\ref{mhfull2}), we obtain the full results for $m_{h,H}^2$. To avoid
a proliferation of scales tangential to the question of decoupling, we
turn off $A_\nu-\mu\cot\beta$ and the $B_\nu$ parameter, and fix a
common scale $m_S$, where $m_{\tilde{L}}=m_{\tilde{R}}\equiv
m_S$.\footnote{The case where $m_{\tilde R}$ and/or $B_\nu$ are
  parametrically larger than the electroweak scale will be briefly
  considered in Section 3.4.}  We expand to first order in
$m^2/m_M^2$, where $m\in\{m_Z,m_S,m_D\}$, and to leading order in
powers of $m_Z$, which is the smallest mass scale when the
superpartner masses, the CP-odd Higgs mass $m_A$, and the Dirac mass
are large.  Note that keeping only the leading order in $m_Z$ is
equivalent to taking $\alpha\simeq\beta-\pi/2$ (since
the vev $v$ aligns with the light state $h$ in this limit).
At leading-logarithmic order, we find that the lightest
Higgs mass squared is shifted relative to its tree level value in the
two renormalization schemes by an amount
\begin{align}
\bigg(\Delta m_h^2\bigg)_{\rm DEC}&\simeq\frac{g^2 m_Z^2}{48\pi^2 c_W^2}\cos^2 2\beta\log\frac{m_S}{m_Z}-\frac{g^2m_D^4m_S^2}{4\pi^2c_W^2m_M^2m_Z^2} \log\frac{m_M}{m_S}\;,\nonumber\\[8pt]
\bigg(\Delta m_h^2\bigg)_{{\rm HM}}&\simeq\frac{g^2m_Z^2}{48\pi^2 c_W^2}\log\frac{m_S}{m_Z}-\frac{g^2m_D^4m_S^2}{4\pi^2c_W^2m_M^2m_Z^2\sin^2\beta} \log\frac{m_M}{m_S}\;,
\label{eq:mhresult}
\end{align}
where $c_W\equiv\cos\theta_W=m_W/m_Z$.

The first terms on the right-hand side of Eq.~(\ref{eq:mhresult}) are
contributions from left-handed sneutrino loops and are insensitive to
the heavy right-handed neutrino scale. These terms also appear in the
ordinary MSSM without neutrino masses. For TeV-scale superpartners,
these terms shift the Higgs mass by 100--200 MeV. The second terms are
leading corrections from the Majorana sector and decouple rapidly as
$\log {m_M}/m_M^2$, giving shifts that are generically less than a
billionth of an eV.  Including corrections of $\mathcal{O}(m_Z^2/m_A^2)$
is equivalent to keeping the tree-level mixing parameter $\alpha$ as
a free parameter.  In this case, the expressions given in
Eq.~(\ref{eq:mhresult}) are somewhat more complicated (with non-trivial
$\alpha$-dependence), but the structure of these results are
maintained.  Contributions that would be sensitive to the
physics of the right-handed neutrino sector would yield additional
terms in Eq.~(\ref{eq:mhresult}) of $\mathcal{O}(m_D^2)$.  However,
using the explicit expressions given in Appendix A, it is
straightforward to verify that such terms \textit{exactly cancel}
in both the HM and DEC schemes, independently of the value of $\alpha$.

The decoupling behavior exhibited in Eq.~(\ref{eq:mhresult})
depends on how the light neutrino masses are
allowed to change as $m_M$ is taken large. Since the overall scale of
the light neutrino masses is not known, $m_D$ can be held fixed while
$m_M$ is increased, in which case both the light neutrino masses and
the second terms in Eq.~(\ref{eq:mhresult}) strictly decrease. On the
other hand, one could also hold the light neutrino mass scale
fixed. In this case, because of the seesaw mechanism present in
Eq.~(\ref{eq:mnu}), the second terms in Eq.~(\ref{eq:mhresult}) are
proportional to $m_\nu^2$ and lose their $m_M^{-2}$ decoupling
behavior.  Of course, this loss of decoupling is illusory, as the
$m_M^{-2}$ behavior is hidden inside $m_\nu^2$ via the seesaw relation
$m_\nu\sim m_D^2/m_M$.  Under the assumption that
$y_\nu\lsim\mathcal{O}(1)$, it follows that $m_D$ cannot be larger
than the electroweak scale, in which case $m_\nu$ is at most of order
$1$ eV for a suitably chosen right-handed neutrino mass scale.  Hence,
the magnitude of the corrections to $m_h$ due to the right-handed
neutrino sector are always minuscule.

For the calculation of $\Delta m_h^2$ in the HM scheme, we avoided the direct computation of $\delta\tan\beta$ by taking advantage of the sum rule, substituting everywhere the tree level expression for $m_H^2$. Therefore, as a check of Eq.~(\ref{eq:mhresult}), we can compute the relation between $\tan\beta$ in the two schemes and see if it is consistent with the difference in the two computations of $\Delta m_h^2$.

The relation between the renormalized $\tan\beta$ parameters is determined by the counterterms,
\begin{align}
\tan\beta_{\rm HM}=\tan\beta_{\rm DEC}+\delta\tan\beta_{\rm HM}-\delta\tan\beta_{\rm DEC}\;,
\label{eq:tbschemes}
\end{align}
where $\delta\tan\beta_{\rm DEC}$ is given by Eq.~(\ref{dtanbdec}) and $\delta\tan\beta_{\rm HM}$ is given by Eq.~(\ref{deltatbHM}).
Hence, the shift in the one-loop prediction for $m_h^2$ incurred by changing schemes is given by inserting Eq.~(\ref{eq:tbschemes}) into the tree level formula for $m_h^2$:
\begin{align}
\bigg(\Delta m_h^2\bigg)_{\rm DEC}-\bigg(\Delta m_h^2\bigg)_{{\rm HM}}\simeq -2m_Z^2\cos^2\beta\sin 4\beta\Bigl[\delta\tan\beta_{\rm HM}-\delta\tan\beta_{\rm DEC}\Bigr]\;.
\label{eq:deltamhschemes}
\end{align}
We find, in the approximations used above for $\Delta m_h^2$,
\begin{align}
\delta\tan\beta_{\rm HM}-\delta\tan\beta_{\rm DEC}\simeq\frac{\tan\beta}{\cos 2\beta}\left(\frac{g^2}{96\pi^2c_W^2}\log\frac{m_S}{m_Z}-\frac{g^2m_D^4m_S^2}{32\pi^2c_W^2m_M^2m_Z^4\sin^4\beta}\log\frac{m_M}{m_S}\right)\;.
\label{eq:deltatb}
\end{align}
It is straightforward to check that inserting Eq.~(\ref{eq:deltatb}) into Eq.~(\ref{eq:deltamhschemes}), the scheme difference obtained in Eq.~(\ref{eq:mhresult}) is recovered.

In non-decoupling subtraction schemes such as $\overline{\rm DR}$,
the non-decoupling contributions to the one-loop corrected Higgs mass
given in Eq.~(\ref{mhfull}) enter via the $\tan\beta$ counterterm.
Using the results of Eqs.~(\ref{dtbhm}) and (\ref{dtbdr}) given in Appendix~A,
\begin{align}
\label{eq:DRDEC}
\delta\tan\beta_{\rm DEC}-\delta\tan\beta_{\rm \overline{DR}}\simeq
\frac{g^2m_D^2}{32\pi^2c_W^2m_Z^2\sin 2\beta}\left(1-\log\frac{m_M^2}{Q^2}\right)\;,
\end{align}
where $Q$ is the renormalization scale.
As noted in Ref.~\cite{Heinemeyer:2010eg}, the partial
decoupling-by-hand of the ${{\rm m}\overline{\rm DR}}$ scheme can be achieved in the
$\overline{\rm DR}$ scheme by taking $Q^2=m_M^2$. However, a finite
non-logarithmic term remains that also must be subtracted by hand if
$\tan\beta$ is to be a genuine low-energy parameter that can be
determined from experimental measurements far below the seesaw
scale. Indeed, one could simply extend the ${{\rm m}\overline{\rm
    DR}}$ scheme by performing this extra subtraction. The end result
is equivalent to the DEC scheme at leading order in our expansions.

To make further contact with the results of
Ref.~\cite{Heinemeyer:2010eg},
we first note that Eq.~(\ref{mhfull}) can be rewritten as
\begin{align} \label{mhfullalt1}
m_h^2=m_{ht}^2 - \widehat{\Sigma}_{hh}(m_{ht}^2)\;,
\end{align}
where $\widehat{\Sigma}_{hh}(p^2)$ is defined in Eq.~(3.7a) of
Ref.~\cite{Heinemeyer:2010eg}.\footnote{Note that the self-energy and tadpole
  functions in the conventions of Ref.~\cite{Heinemeyer:2010eg} differ by
  an overall sign from those defined in this paper.
This is the origin of the minus sign in Eq.~(\ref{mhfullalt1}).} If the
two-loop contributions generated by products of self-energy
functions are neglected
in Eq.~(3.2) of Ref.~\cite{Heinemeyer:2010eg}, then the pole in the
matrix propagator corresponding to the light CP-even Higgs mass is given by
\begin{align} \label{mhfullalt2}
m_h^2=m_{ht}^2 - \widehat{\Sigma}_{hh}(m_h^2)\;,
\end{align}
where $m_h^2$ appearing on the right-hand side above is the one-loop corrected Higgs mass. 
Note that the fact that the argument of $\widehat\Sigma_{hh}$ is $m_h^2$ rather
than $m_{ht}^2$ means that partial two-loop information is being included in the expression for
the one-loop corrected Higgs mass.  In this case, Eq.~(3.7a) of Ref.~\cite{Heinemeyer:2010eg}
implies that the loop-corrected Higgs mass given by Eq.~(\ref{mhfullalt2})
is equivalent to Eq.~(\ref{mhfull}) with the following replacement,
\begin{align}
\Sigma_{hh}(m_{ht}^2)\rightarrow \bigl[\Sigma_{hh}(p^2)-\delta
\mathcal{Z}_{hh} (p^2-m_{ht}^2)\bigr]\biggl|_{p^2=m_h^2},
\label{mhfullmod}
\end{align}
where [cf.~Eq.~(3.10a) of Ref.~\cite{Heinemeyer:2010eg}],
\begin{align} \label{zhh}
\delta\mathcal{Z}_{hh}=\sin^2\alpha~\delta\mathcal{Z}_{H_d} + \cos^2\alpha~\delta\mathcal{Z}_{H_u}\;.
\end{align}

We now examine in more detail how decoupling occurs in the expression for the loop-corrected Higgs mass.
It is convenient to define a momentum-dependent Higgs squared-mass,
\begin{align} 
m_h^2(p^2) &\equiv  m_{ht}^2-\widehat\Sigma_{hh}(p^2)\nonumber \\
&= m_h^2(m_{ht}^2)+\Sigma_{hh}(p^2)-\Sigma_{hh}(m_{ht}^2)-\delta\mathcal{Z}_{hh}(p^2-m_{ht}^2)\nonumber \\
&\equiv \overline{m}_h^2(p^2)-\delta\mathcal{Z}_{hh}(p^2-m_{ht}^2)\,,\label{mhfullp}
\end{align}
where $\overline{m}_h^2(p^2)$ corresponds to the result of Eq.~(\ref{mhfull}) after replacing $\Sigma_{hh}(m_{ht}^2)$
with $\Sigma_{hh}(p^2)$.  By choosing either $p^2=m_{ht}^2$ or $p^2=m_{h}^2$, we recover either 
Eq.~(\ref{mhfullalt1}) or Eq.~(\ref{mhfullalt2}), respectively.
The potential non-decoupling behavior lies in the $\mathcal{O}(m_D^2)$ contributions to the loop-corrected Higgs mass.
In Appendix~A, we give the leading terms
contributing at $\mathcal{O}(m_D^2)$ in the individual self-energy
functions, tadpoles, and the $\tan\beta$ counterterm.
None of the individual terms that appear in the expression for the loop-corrected Higgs mass vanish in the large $m_M$ limit.
However, given a decoupling scheme for $\delta\tan\beta$ [and $\delta\mathcal{Z}_{hh}$, if Eq.~(\ref{mhfullmod}) is used],
then the non-decoupling terms cancel exactly in the Higgs mass prediction,
leaving only $m_M^2$-suppressed terms at $\mathcal{O}(m_D^4)$.

It is instructive to evaluate the $\mathcal{O}(m_D^2)$ contributions to $m_h^2(p^2)$ in the DEC scheme.  Using the
results of Appendix A, we readily find that
\begin{align}
\overline{m}_h^2(p^2)\biggl|_{\mathcal{O}(m_D^2)} & =
-\frac{g^2m_D^2}{64\pi^2c_W^2m_Z^2\sin^2\beta}\bigg(\frac{1}{\epsilon}-\gamma+\log4\pi+1-\log\frac{m_M^2}{Q^2}\bigg)\nonumber\\[6pt]
&\qquad\quad\times\biggl[p^2-m_A^2+m_{Ht}^2-m_Z^2+\cos2\beta(m_Z^2-m_A^2)+\cos2\alpha(p^2-m_{Ht}^2)\biggr]\;,
\label{eq:nondecHM}
\end{align}
where the pole at $\epsilon=0$ indicates that the ultraviolet divergences have not yet canceled [cf.~Eq.~(\ref{Qtilde}) of Appendix A].
We can simplify Eq.~(\ref{eq:nondecHM}) by using the tree-level sum
rule $m_{ht}^2=m_A^2-m_{Ht}^2+m_Z^2$ and the tree-level mixing angle relation
\begin{align} \label{mixang}
\cos2\alpha(m_{Ht}^2-m_{ht}^2)=\cos2\beta(m_Z^2-m_A^2)\;.
\end{align}
The end result is
\begin{align}
\overline{m}_h^2(p^2)\biggl|_{\mathcal{O}(m_D^2)}=-\frac{g^2m_D^2\cos^2\alpha}{32\pi^2c_W^2m_Z^2\sin^2\beta}(p^2-m_{ht}^2)\left(\frac{1}{\epsilon}-\gamma+\log4\pi+1-\log\frac{m_M^2}{Q^2}\right)\;.
\label{eq:nondecmDEC}
\end{align}

To complete the computation of $m_h^2(p^2)$, we make use of Eqs.~(\ref{deltaZs}) and (\ref{zhh}) and the $\mathcal{O}(m_D^2)$ expressions given in
Eqs.~(\ref{sighhp}) and (\ref{sigHHp}),
\begin{align} \label{zhhdec}
\delta\mathcal{Z}_{hh}\biggl|_{\mathcal{O}(m_D^2)}=-\frac{g^2m_D^2\cos^2\alpha}{32\pi^2c_W^2m_Z^2\sin^2\beta}\left(\frac{1}{\epsilon}-\gamma+\log4\pi+1-\log\frac{m_M^2}{Q^2}\right)\;.
\end{align}
Using Eq.~(\ref{mhfullp}), it follows that the $\mathcal{O}(m_D^2)$ contributions to $m_h^2(p^2)$ exactly cancel in the DEC scheme.  
This decoupling has already been demonstrated for the one-loop corrected Higgs mass defined by Eq.~(\ref{mhfull}) in the DEC scheme
[cf.~Eq.~(\ref{eq:mhresult})].

One can repeat the above calculation in the HM scheme, where $\overline{m}_h^2(p^2)$ 
is most easily obtained using Eq.~(\ref{eq:mhsumrule}), which yields
\begin{equation}
\overline{m}_h^2(p^2)=m_{ht}^2+\Sigma_{hh}(p^2)+\Sigma_{HH}(m_H^2)-\Sigma_{ZZ}(m_Z^2)-\Sigma_{AA}(m_A^2)-\Sigma_{GG}(0)\,.
\end{equation}
Evaluating the self-energy functions using the results of Appendix A, we again recover the result of
Eq.~(\ref{eq:nondecHM}).  For $p^2=m_{ht}^2$, the $\mathcal{O}(m_D^2)$ terms vanish exactly and the decoupling behavior is
established, as previously demonstrated.
In the case of $p^2\neq m_{ht}^2$, we need a separate definition of the Higgs wave function counterterms.
Here, the natural choice is an on-shell scheme, which fixes the residues of the corresponding pole masses to unity.
In this scheme, the $\mathcal{O}(m_D^2)$ contributions to $\delta\mathcal{Z}_{hh}\bigl|_{\mathcal{O}(m_D^2)}$ are the same
as those of the DEC scheme, since the $\mathcal{O}(m_D^2)$ contributions to $d\Sigma_{hh}(p^2)/dp^2$ and $d\Sigma_{HH}(p^2)/dp^2$
are independent of $p^2$.  Thus, it again follows that the $\mathcal{O}(m_D^2)$ contributions to $m_h^2(p^2)$ exactly cancel in the HM scheme.

In contrast, consider the computation of $m_h^2(p^2)$ in the $\overline{\rm DR}$ scheme.  Due to the modification of the $\tan\beta$ counterterm 
[cf.~Eq.~(\ref{eq:DRDEC})], an extra term is obtained in the evaluation of $m_h^2(m_{ht}^2)$ [cf.~Eq.~(\ref{mhfull})].
It follows that in the $\overline{\rm DR}$ scheme,
\begin{align}
\overline{m}_h^2(p^2)\biggl|_{\mathcal{O}(m_D^2)}&=\frac{g^2m_D^2}{32\pi^2c_W^2}\biggl\{\cot\beta\sin\bigl(2(\beta+\alpha)\bigr)\left(1-\log\frac{m_M^2}{Q^2}\right) \nonumber \\
&\qquad\qquad\qquad -\frac{\cos^2\alpha}{\sin^2\beta}\left(\frac{p^2-m_{ht}^2}{m_Z^2}\right)\left(\frac{1}{\epsilon}-\gamma+\log4\pi+1-\log\frac{m_M^2}{Q^2}\right)\biggr\}\;.
\label{eq:nondecmDR}
\end{align}
To obtain the corresponding $\overline{\rm DR}$ expression for
$\delta\mathcal{Z}_{hh}\bigl|_{\mathcal{O}(m_D^2)}$, we retain 
$\epsilon^{-1}-\gamma+\log 4\pi$  in Eq.~(\ref{zhhdec})
and discard the remaining terms.   Thus in the $\overline{\rm DR}$ scheme, Eq.~(\ref{mhfullp}) yields
\begin{align}
{m}_h^2(p^2)\biggl|_{\mathcal{O}(m_D^2)}=\frac{g^2m_D^2}{32\pi^2c_W^2}\left[\cot\beta\sin\bigl(2(\beta+\alpha)\bigr)
-\frac{\cos^2\alpha}{\sin^2\beta}\left(\frac{p^2-m_{ht}^2}{m_Z^2}\right)\right]\left(1-\log\frac{m_M^2}{Q^2}\right)\,.
\label{eq:nondecoupDR}
\end{align}
In the $\rm{m}\overline{\rm DR}$ scheme of Ref.~\cite{Heinemeyer:2010eg}, one sets $Q^2=m_M^2$ to remove the logarithm, but the constant term
remains and decoupling is not satisfied.  The loop-corrected Higgs mass advocated in Ref.~\cite{Heinemeyer:2010eg} corresponds to setting $p^2=m_h^2$
in $m^2_h(p^2)$ [cf.~Eq.~(\ref{mhfullalt2})].
In this case, there are two separate contributions to the non-decoupling behavior, corresponding to the two terms obtained in Eq.~(\ref{eq:nondecoupDR}).
In the $\rm{m}\overline{\rm DR}$  scheme, the second term of Eq.~(\ref{eq:nondecoupDR}) is negative and provides the dominant source of the
Higgs mass shift at large $\tan\beta$.  Indeed, it is of the correct order of magnitude to explain the decrement in $m_h$ obtained in the numerical analysis of Ref.~\cite{Heinemeyer:2010eg}.

Thus, we have located the sources of the non-decoupling behavior found in Ref.~\cite{Heinemeyer:2010eg}.   However, we note that even in a
consistent one-loop truncation where $p^2=m_{ht}^2$ is taken to evaluate the loop-corrected Higgs mass, there is still a residual 
non-decoupling behavior in the $\rm{m}\overline{\rm DR}$ scheme, which enters via the $\tan\beta$ counterterm (which fixes the definition of $\tan\beta$).  
In contrast, by employing a decoupling scheme to fix the $\tan\beta$ counterterm (and the Higgs wave function counterterms if separately needed),
one is guaranteed a loop-corrected Higgs mass that is completely insensitive to the physics at the right-handed neutrino scale (assuming this scale
lies significantly above the SUSY-breaking scale).

\subsection{Effective Field Theory Estimates of the Higgs Mass Shift}
In Ref.~\cite{Heinemeyer:2010eg} it was argued that large corrections to $m_h$ could be traced to terms proportional to the external momenta in the self-energy functions.
Such terms would not appear in the usual effective potential calculation. However, we have found that in a consistent one-loop truncation, such large corrections do not appear in the full expression for the physical Higgs mass when expressed in terms of parameters that can be measured directly in the low-energy effective theory.
Therefore, it should be possible to derive the parametric properties of the leading terms presented in Section 3.1 directly from corrections to the Higgs quartic coupling in the effective potential, as computed in effective field theory (EFT)---the natural framework for dealing with large mass hierarchies. For simplicity, we will work primarily in the small-$m_Z$ limit, where the vev $v$ aligns with the light state $h$ such that $\alpha\rightarrow\beta-\pi/2$.

The $m_Z^2$ term we found in $\Delta m_h^2$ is just the usual contribution at low scales from the $D$-term coupling $|H_u|^2|\tilde{L}^2$, and is insensitive to the $m_M$ threshold. What about the subleading term? Imagine that we integrate out the right-handed neutrino and sneutrino at the right-handed neutrino mass threshold.  Above this scale, the running of $\lambda$ (the coefficient of the quartic self-coupling $\frac{1}{8}h^4$ in the effective Lagrangian) is supersymmetric, but the TeV-scale soft mass splits the scalar and fermion states, leading to a logarithmic correction to $\lambda$ from the right-handed sneutrino bubble diagram:
\begin{align}
\Delta m_h^2 =2(\Delta \lambda) v^2 \sim \frac{m_D^4}{v^2}\log\frac{m_{\tilde{N}}^2}{m_N^2}\sim \frac{m_D^4m_S^2}{v^2m_M^2}\;.
\end{align}
This term is certainly present in the corrections, but it is $m_M$-suppressed and has no log enhancement, so it is not the source of the second terms in Eq.~(\ref{eq:mhresult}). In addition to direct contributions to $\lambda$, we also generate an approximately supersymmetric higher-dimensional coupling,
\begin{align}
\Delta W= \frac{y_\nu^2}{m_M}LH_uLH_u\;.
\end{align}
This coupling affects the running of $\lambda$ when supersymmetry is
broken via the diagrams in Fig.~1. The dominant contribution
comes from the sneutrino diagram,
\begin{align}
\frac{\partial\lambda}{\partial\log Q^2}\approx \frac{y_\nu^4 m_S^2 \sin^4\beta}{8\pi^2 m_M^2}\;.
\end{align}
 Running the quartic coupling down from $m_M$ to $m_S$ and recalling that $v=\sqrt{2} m_W/g$, we obtain at leading logarithmic order,
 \begin{align}
\Delta m_h^2 = -\frac{m_D^4m_S^2}{2\pi^2v^2m_M^2}\log\frac{m_M}{m_S}\;,
\label{eq:DECEFT}
\end{align}
matching the terms in Eq.~(\ref{eq:mhresult}) in the DEC scheme.

\begin{figure}[t]
\begin{center}
\hspace*{-0.75cm}
\includegraphics[width=1\textwidth,clip=true, viewport=0.5in 8in 7.5in 10.5in]{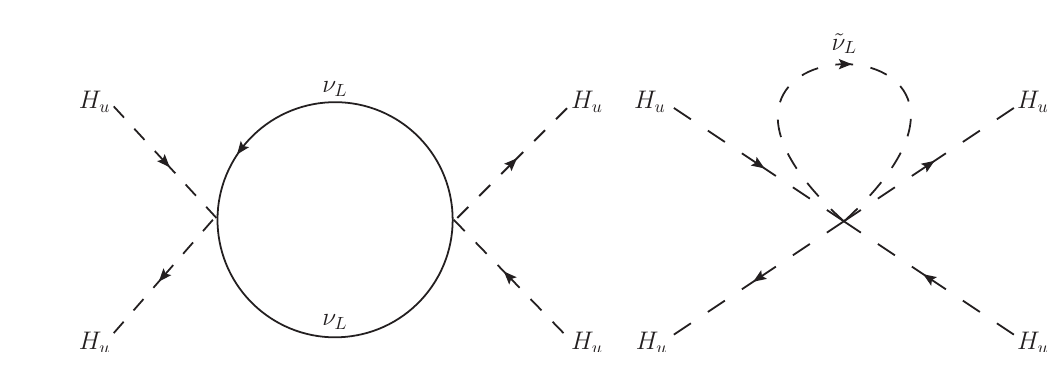} \hspace{3mm}
\vspace{0.2cm}
\caption{Diagrams contributing to the running of the Higgs quartic below the right-handed neutrino mass scale.}
\end{center}
\end{figure}

To understand why we obtained the DEC scheme result instead of the HM scheme result, and how the latter can be reproduced, we have to consider the definition of $\tan\beta$ in the effective theory. Up to threshold corrections that are subleading (not log-enhanced), $\tan\beta\lsub{\rm EFT}=\tan\beta\lsub{\rm full}$ at the matching scale $Q=m_M$. Therefore, the tree-level boundary condition for the Higgs self-coupling $\lambda$ takes the usual form,
\begin{align}
\lambda=\tfrac{1}{4}(g_1^2+g_2^2)\cos^2 2\beta\lsub{\rm EFT}\,,
\end{align}
at the matching scale. To obtain the $m_M$-dependent corrections to a low-energy prediction for $m_h$, we should include not only the shift of Eq.~(\ref{eq:DECEFT}), but also contributions obtained by rewriting $\cos^2 2\beta$ in the EFT at $Q=m_M$ in terms of $\cos^2 2\beta$ in the EFT at $Q=m_S$.

Below $m_M$, the dimension-5 operator contributes to the running of $\tan\beta$ in a scheme-dependent way. It is straightforward to check that the beta-function for $\tan\beta$ does not contain terms proportional to $m_S^2/m_M^2$
in the DEC scheme or any minimal subtraction scheme, where the field-strength renormalization counterterms are set by derivatives of self-energies with respect to $p^2$.  The relevant diagrams are obtained by setting two external legs to $v_u$ in Fig.~1, which makes it clear that the sneutrino loop is independent of $p^2$. Therefore, in the DEC scheme, the corrections to $m_h^2$ from the running of $\tan\beta$ are higher-order in the $m_Z$ expansion, and are not required to reproduce Eq.~(\ref{eq:mhresult}).

In contrast, the $\tan\beta$ counterterm in the HM scheme is controlled by the self-energies themselves instead of their $p^2$ derivatives. At leading order in the $m_Z$ expansion, Eq.~(\ref{deltatbHM}) with $\alpha=\beta-\pi/2$ yields:
\begin{align}
(\delta\tan\beta)_{\rm HM}=-\frac{1}{2m_Z^2\cos^2\beta\sin4\beta}\bigl[\Sigma_{AA}(m_A^2)-\Sigma_{HH}(m_H^2)\bigr]\;.
\end{align}
Therefore, the sneutrino contributions to $\Sigma_{HH}(m_H^2)$ and $\Sigma_{AA}(m_A^2)$ can provide $m_S^2/m_M^2$ terms in the running of $\tan\beta$. Explicitly,
\begin{align}
\frac{\partial (\tan\beta)_{\rm HM}}{\partial\log Q^2}= \frac{1}{2m_Z^2\cos^2\beta\sin4\beta}\frac{y_\nu^4m_S^2v_u^2\cos^2\beta}{4\pi^2m_M^2}\;,
\end{align}
which at leading-log yields,
\begin{align}
\Delta m_h^2 &= 2m_Z^2\cos^2\beta\sin4\beta\frac{\partial (\tan\beta)_{\rm HM}}{\partial\log Q^2}\log\frac{m_S^2}{m_M^2}\nonumber\\[6pt]
&=-\frac{m_D^4m_S^2}{2\pi^2v^2m_M^2\tan^2\beta}\log\frac{m_M}{m_S}\;.
\label{eq:tbHMrun}
\end{align}
Adding Eq.~(\ref{eq:tbHMrun}) to Eq.~(\ref{eq:DECEFT}), we recover the leading HM scheme expression given by the full theory in Eq.~(\ref{eq:mhresult}).

A more complete effective field theory analysis of the threshold corrections from the right-handed neutrino/sneutrino sector is beyond the scope of this paper. However, our full-theory calculation makes clear how decoupling will manifest at the thresholds. Loop diagrams involving right-handed neutrinos or sneutrinos will indeed provide non-decoupling finite contributions to the low-energy effective Higgs self-coupling $\lambda$ during matching, but these contributions will be absorbed by finite and unobservable shifts in $\tan\beta$.

\subsection{Numerical Results}

\begin{figure}
\begin{center}
\hspace*{-0.75cm}
\includegraphics[width=0.48\textwidth]{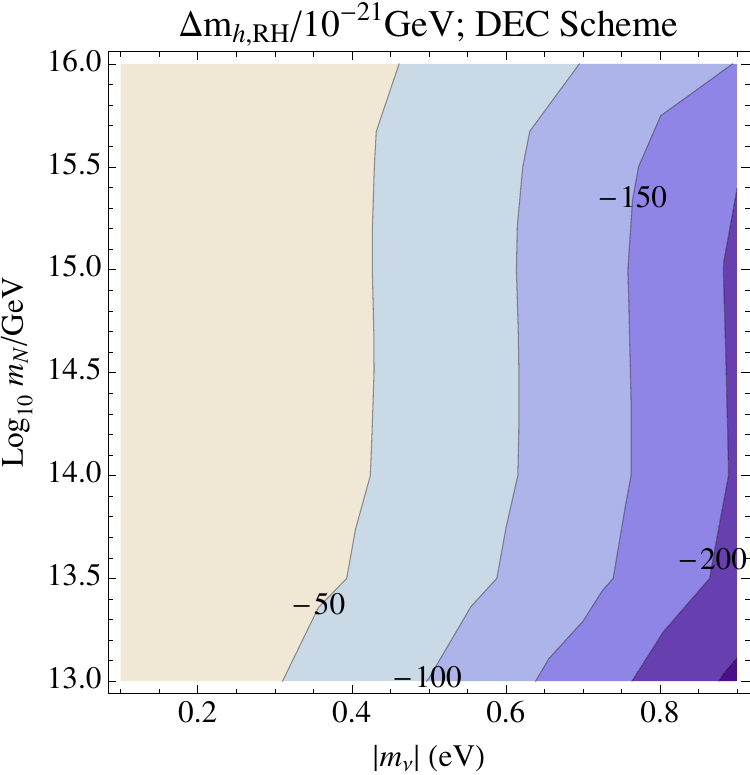} \hspace{3mm}
\includegraphics[width=0.48\textwidth]{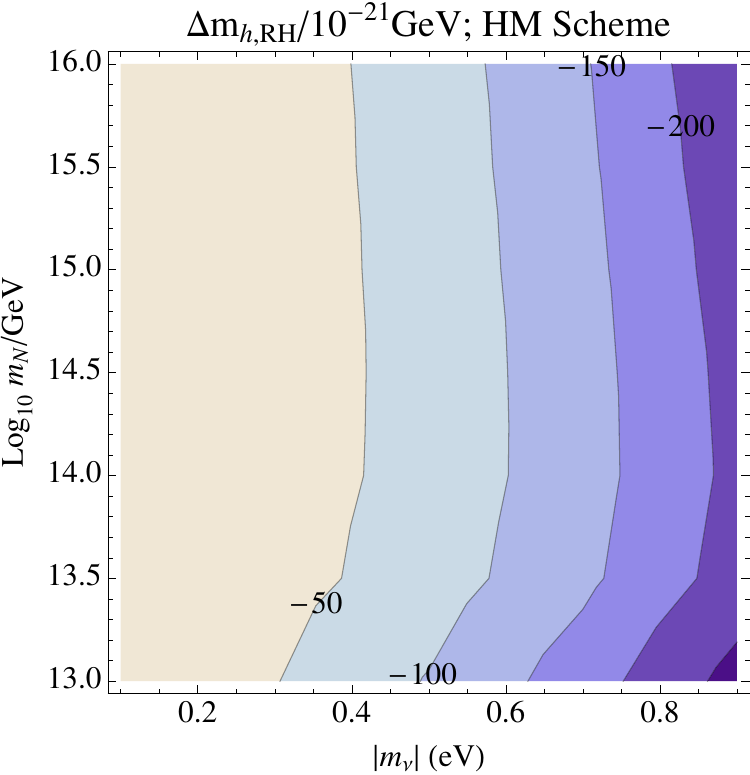} \hspace{3mm}
\caption{Left panel: The shift in the Higgs mass due to right-handed (s)neutrinos in the decoupling (DEC) scheme at different points on the neutrino mass plane, for values of the parameters given in the text.  Right panel: The same in the HiggsMass (HM) scheme. $\Delta m_{h,RH}$ is defined in Eq.~(\ref{eq:deltamhdef}).}
\label{fig:fullnumerics}
\end{center}
\end{figure}

The full one-loop analytic formulae for the Higgs mass shifts in the decoupling schemes are too complicated to reproduce here. On the other hand, the approximations used above do not rule out the possibility of large corrections proportional to $m_M^2$ or $\log m_M^2$ appearing at higher order in the $m_Z$ expansion or in non-logarithmic terms. To demonstrate that such terms are not present, we have numerically evaluated
the full one-loop (s)neutrino contribution to $m_h$ as a function of $|m_\nu|$ and $m_N$, with the pure left-handed sneutrino contribution subtracted out. For definiteness, we define
\begin{align}
\Delta m_{h,RH}\equiv \sqrt{m_{ht}^2+\Delta m^2_{h,RH}}-m_{ht}\;,
\label{eq:deltamhdef}
\end{align}
which can be thought of as an upper bound on the contribution to $m_h$ from the right-handed (RH) sector.  The results are exhibited in Fig.~\ref{fig:fullnumerics}.

If additional sectors are included to raise $m_h$ from $m_{ht}\sim m_Z$ to 126 GeV, $\Delta m_{h,RH}$ will be further suppressed by about 40\%, although this is clearly unimportant in light of the overall scale of the corrections in Fig.~\ref{fig:fullnumerics}.  Other parameters in the figure are fixed to the values $A_\nu=B_\nu=m_S=1$ TeV, $\mu=200$ GeV, and $\tan\beta=5$. As $m_M$ is increased for fixed $m_D$, we move towards the upper-left corner of the plot, where the mass shift is minimal: this trend establishes decoupling in the fixed $m_D$-sense. If we increase $m_M$ and $m_D$ so that the light physical neutrino mass $m_\nu$ is fixed, we see that the corrections are roughly constant, also as expected. In either case the overall magnitude of the corrections is never larger than about
$10^{-10}$ eV, which is consistent with our estimate from Eq.~(\ref{eq:mhresult}).

\subsection{Large SUSY-Breaking in the Right-Handed Sector}

Consider the impact of choosing values for the SUSY-breaking parameters $m_{\tilde{R}}^2$ and $B_\nu$ that are large compared to the other SUSY-breaking parameters.
If soft squared-mass parameter $m_{\tilde{R}}^2$ becomes of order $m_M^2$, then the contribution to the Higgs quartic coupling from the running between $m_{\tilde{N}}^2$ and $m_N^2$ no longer decouples with large $m_M$. The Higgs mass receives a correction of order
\begin{align}
\Delta m_h^2\sim \frac{m_D^4}{v^2}\log\left(\frac{m_M^2+m_{\tilde{R}}^2}{m_M^2}\right)\;,
\end{align}
in complete analogy to the contribution from the top squarks. However, $m_{\tilde{R}}^2$ also enters into the one-loop RGE for the Higgs mass parameter $m_{H_u}^2$, and therefore exacerbates the little hierarchy problem when $m_{\tilde{R}}^2\gg m_Z^2$. For this reason it is preferable to keep $m_{\tilde{R}}^2$ of the same order as other squark and slepton squared-mass parameters.

When the soft mass parameter $B_\nu$ dominates the SUSY-breaking parameters, it splits the $CP$-even and $CP$-odd right-handed sneutrinos according to $m_{\tilde{N}_{\pm}}\approx m_M\pm B_\nu$. It also alters the running of the Higgs quartic coupling at high energy scales and inhibits decoupling. Running between $m_{\tilde{N}_+}^2$ and $m_{\tilde{N}_-}^2$ yields a correction to the Higgs mass of order
\begin{align}
\Delta m_h^2&\sim \frac{m_D^4}{v^2}\log\left(\frac{m_{\tilde{N}_+}m_{\tilde{N}_-}}{m_N^2}\right)
\simeq\frac{m_D^4}{v^2}\log\left(\frac{m_M^2-B_\nu^2}{m_M^2}\right)\;.
\end{align}
The primary distinction from the case of large $m_{\tilde{R}}^2$ is that $B_\nu$ lowers the geometric mean of the right-handed sneutrino masses, making the logarithm negative and decreasing the Higgs mass. However, as in the case of $m_{\tilde{R}}^2$, there is a good reason to keep $B_\nu\ll m_M$.  In particular, a large value of $B_\nu$ generates a large contribution to $\tilde\nu_L$--$\tilde\nu_R$  mixing, which in turn generates a one-loop correction to the light neutrino masses that swamps the tree-level seesaw contribution if $B_\nu\gtrsim 10^3 m_{\tilde{\nu}_L}$~\cite{Grossman:1997is}.

In both the large $m_{\tilde{R}}^2$ and large $B_\nu$ scenarios, the contribution to $m_h$ from the left-handed sector diagrams of Fig.~1 are subdominant. The large right-handed neutrino-sneutrino mass splittings change the argument of the logarithm, but the contribution remains suppressed by the left-handed neutrino-sneutrino mass splitting controlled by $m_{\tilde{L}}^2$.

\section{Conclusions}
A recent analysis~\cite{Heinemeyer:2010eg} has argued that adding a right-handed neutrino and sneutrino to the MSSM could generate a sizable radiative contribution to the lightest Higgs boson mass in the case of a large right-handed neutrino mass scale, even if all soft SUSY-breaking parameters remain at the TeV scale. Such a non-decoupling effect would cast doubt on the notion that the Higgs mass can be reliably calculated in a weak-scale supersymmetric theory in terms of measurable TeV-scale parameters.
In this paper we have reanalyzed the radiative corrections to the Higgs mass from the right-handed neutrino sector.

In the analysis presented in this work, we began with a review of the
computation of one-loop corrections to the physical masses of the
neutral Higgs bosons of the MSSM, streamlining the derivation, providing compact general formulae for the spectrum, and reviewing the decoupling properties of various $\tan\beta$ renormalization schemes.  In our consideration of the relevance of decoupling, we distinguished two cases. First, we commented briefly on the possibility that $\tan\beta$ cannot be independently measured in any scheme.  For example, this could occur simply because all MSSM degrees of freedom are too heavy, in which case the decoupling properties of the scheme used to define $\tan\beta$ are irrelevant.  However, the corresponding MSSM Higgs mass prediction cannot be tested, and the most that can be achieved is a scheme-dependent constraint on the superpartner mass scale and $\tan\beta$.   Much more relevant for phenomenology is the alternative case, where some MSSM particles with $\tan\beta$-sensitive couplings can be accessed in collider experiments.  In this latter case, one can predict the masses of the MSSM Higgs bosons in terms of quantities that are directly accessible to experimental measurements.  These predicted masses are completely insensitive to
physics at mass scales significantly larger than the scale of SUSY-breaking
(such as the high-scale seesaw sector employed in a theory of neutrino masses).  Consequently, it is especially convenient to define the parameter $\tan\beta$ using a renormalization scheme that respects decoupling, since the
expressions for the MSSM Higgs masses (which depend explicitly on $\tan\beta$) will then manifestly exhibit the expected decoupling behavior.

Applying the general mass formulae to the right-handed neutrino sector, we derived expressions for the leading contributions in two decoupling schemes, and
found that the magnitude of the corrections to the Higgs mass are
utterly negligible. The expected decoupling behavior is observed if
the right-handed neutrino mass scale is taken large while other input
parameters are held fixed. The structure of the leading correction
terms is easily recovered from effective field theory arguments.
Finally, to go beyond the approximate formulae, we performed a
numerical analysis including all contributing one-loop terms. We find
that the corrections remain negligible and are well-reproduced by the
leading terms. Since all the relevant couplings are weak, it is
sufficient to work to one-loop order.  In particular, the effective
field theory analysis gives us confidence that our results will not
change with the inclusion of two-loop and higher-order effects.  Thus,
we conclude that the right-handed neutrino mass scale plays no
significant role in the determination of the Higgs spectrum in
weak-scale supersymmetric models.

\bigskip

\subsection*{Acknowledgments}
We would like to acknowledge fruitful discussions with
Sven Heinemeyer and Maria Herrero
concerning the work of Ref.~\cite{Heinemeyer:2010eg}.  We are also
grateful to the anonymous referee whose critique of earlier versions
of this manuscript resulted in significant improvements to the presentation.
PD and HEH are supported in part by U.S. Department of
Energy grant number DE-FG02-04ER41286.

\bigskip\bigskip
\centerline{\bf \Large APPENDIX}
\vspace{-0.1in}
\appendix
\section{Approximate Renormalized Self-Energies and Tadpoles}

It is convenient to have
analytic approximations for the self-energy functions and
tadpoles in order to see how the terms sensitive
to the seesaw scale explicitly cancel in the expressions for the
Higgs masses
[Eqs.~(\ref{mhfull}) and (\ref{mhfull2})].
Following Ref.~\cite{Heinemeyer:2010eg}, we perform a series
expansion in powers of $m_D^2$. At $\mathcal{O}(m_D^0)$, the
contributions are insensitive to the seesaw scale. At
$\mathcal{O}(m_D^4)$, each self-energy scales as
$m_M^{-2}$, exhibiting decoupling independently, in agreement
with Ref.~\cite{Heinemeyer:2010eg}. In contrast, decoupling occurs
in the $\mathcal{O}(m_D^2)$ terms due to
nontrivial cancellations among the various terms in Eqs.~(\ref{mhfull})
and (\ref{mhfull2}).

Below we give the $\mathcal{O}(m_D^2)$ contributions to the real parts
of the self-energy functions and tadpoles\footnote{Note that the self-energy and tadpole
  functions in the conventions of Ref.~\cite{Heinemeyer:2010eg} differ by
  an overall sign from those defined in this paper.} in $d$-spacetime dimensions using dimensional
regularization, expanded to
leading order with respect to the mass hierarchy
\begin{equation} \label{hierarchy}
\{m_Z^2,p^2,m_A^2,m_H^2\}\ll m_S^2\ll m_M^2\,.
\end{equation}
It is convenient to adopt the
shorthand notation
\begin{equation}
\log \widetilde{Q}^2\equiv\frac{1}{\epsilon}-\gamma+\log(4\pi Q^2)\,,
\label{Qtilde}
\end{equation}
where $Q$ is the renormalization scale,
$\epsilon\equiv 2-\half d$ and $\gamma$ is Euler's constant.
The $\mathcal{O}(m_D^2)$ contribution to $\Sigma_{hh}(p^2)$
at leading order in the mass hierarchy
[cf.~Eq.~(\ref{hierarchy})]
is given by
\begin{align} \label{sighhp}
\Sigma_{hh}(p^2)=& \frac{g^2 m_D^2}{64 \pi^2 c_W^2 m_Z^2}\Biggl\{
\frac{2\cos^2\alpha}{\sin^2\beta}
\bigg[2m_S^2\log \frac{m_M^2}{\widetilde{Q}^2}+(m_Z^2-p^2)\bigg(1-\log \frac{m_M^2}{\widetilde{Q}^2}\bigg)
\bigg]\nonumber\\[6pt]
   &+m_Z^2 \bigg(1-\log \frac{m_M^2}{\widetilde{Q}^2}\bigg)
\biggl[\cos^2\alpha  (4-3
   \cot^2\beta )+2 \sin 2 \alpha  \cot \beta -\sin^2\alpha\biggr]\Biggr\}\,,
\end{align}
where $p$ is the incoming four momentum.  Likewise, $\Sigma_{HH}(p^2)$ is obtained
by making the replacement $\alpha\to\alpha-\half\pi$ in Eq.~(\ref{sighhp}),
\begin{align} \label{sigHHp}
\Sigma_{HH}(p^2)=& \frac{g^2 m_D^2}{64 \pi^2 c_W^2 m_Z^2}\,\Biggl\{
\frac{2\sin^2\alpha}{\sin^2\beta}
\bigg[2m_S^2\log \frac{m_M^2}{\widetilde{Q}^2}+(m_Z^2-p^2)\bigg(1-\log \frac{m_M^2}{\widetilde{Q}^2}\bigg)
\bigg]\nonumber\\[6pt]
   &+m_Z^2 \bigg(1-\log \frac{m_M^2}{\widetilde{Q}^2}\bigg)
\biggl[\sin^2\alpha  (4-3
   \cot^2\beta )-2 \sin 2 \alpha  \cot \beta -\cos^2\alpha\biggr]\Biggr\}\,.
\end{align}

For completeness, we provide the $\mathcal{O}(m_D^2)$ contribution to the
real parts of all the other relevant
self-energy functions [at leading order in the mass hierarchy,
Eq.~(\ref{hierarchy})],
\begin{align}
\Sigma_{ZZ}(m_Z^2)= &\frac{g^2 m_D^2}{64 \pi^2 c_W^2 m_Z^2}\,2m_Z^2\bigg(1-\log \frac{m_M^2}{\widetilde{Q}^2}\bigg),\nonumber\\[8pt]
\Sigma_{AA}(m_A^2)=&\frac{g^2 m_D^2}{64 \pi^2 c_W^2 m_Z^2}\Biggl\{\frac{\cos 2 \beta}{\sin^2\beta}
\bigg[2 m_S^2 \log \frac{m_M^2}{\widetilde{Q}^2}-(m_A^2 +m_Z^2)\bigg(1-\log\frac{m_M^2}{\widetilde{Q}^2}\bigg)\bigg]\nonumber\\[6pt]
   &\qquad\qquad\qquad +\frac{1}{\sin^2\beta}\left[2 m_S^2\log \frac{m_M^2}{\widetilde{Q}^2}-m_A^2 \bigg(1-\log \frac{m_M^2}{\widetilde{Q}^2}\bigg)\right]+2 \cos2 \beta\,  m_Z^2 \bigg(1-\log\frac{m_M^2}{\widetilde{Q}^2}\bigg)\Biggr\}\nonumber,\\[8pt]
\Sigma_{GG}(0)=&~\frac{g^2 m_D^2}{64 \pi^2 c_W^2 m_Z^2}\left\{ 4 m_S^2 \log \frac{m_M^2}{\widetilde{Q}^2}-2 \cos 2 \beta\,  m_Z^2 \bigg(1-\log\frac{m_M^2}{\widetilde{Q}^2}\bigg)\right\},\nonumber\\[8pt]
\frac{A_h}{\sqrt{2}\,v}=&\frac{g^2 m_D^2}{64 \pi^2 c_W^2 m_Z^2}\left\{ \frac{\cos \alpha}{\sin\beta}  \Bigg[4 m_S^2 \log \frac{m_M^2}{\widetilde{Q}^2}-m_Z^2
   \bigg(1-\log \frac{m_M^2}{\widetilde{Q}^2}\bigg)\Bigg]\right.\nonumber \\[6pt]
& \qquad\qquad\qquad \left. +m_Z^2 (\sin \alpha  \cos \beta +3 \cos \alpha
   \sin \beta ) \bigg(1-\log \frac{m_M^2}{\widetilde{Q}^2}\bigg)\right\},\nonumber\\[8pt]
\frac{A_H}{\sqrt{2}\,v}=&~\frac{g^2 m_D^2}{64 \pi^2 c_W^2 m_Z^2}\left\{\frac{\sin \alpha}{\sin\beta}  \Bigg[4 m_S^2 \log \frac{m_M^2}{\widetilde{Q}^2}-m_Z^2
   \bigg(1-\log \frac{m_M^2}{\widetilde{Q}^2}\bigg)\Bigg]\right.\nonumber \\[6pt]
& \qquad\qquad\qquad \left. +m_Z^2 (3 \sin \alpha  \sin \beta -\cos \alpha
   \cos \beta ) \bigg(1-\log \frac{m_M^2}{\widetilde{Q}^2}\bigg)\right\}.\label{expansions}
\end{align}

Next, we compute
the $\mathcal{O}(m_D^2)$ contributions [at leading order in the mass hierarchy,
Eq.~(\ref{hierarchy})] to the counterterm
$\delta\tan\beta$ in the various renormalization schemes.
In the HM scheme, $\delta\tan\beta$ is given by Eq.~(\ref{deltatbHM}).
Using the above expressions for the self-energy functions, along
with Eq.~(\ref{AZ}) and the following
tree-level relations (cf.~Eq.~(A.20) of Ref.~\cite{Gunion:1986nh}),
\begin{equation}
m_{ht}^2=-\frac{m_Z^2\cos 2\beta\sin(\beta+\alpha)}{\sin(\beta-\alpha)}\,,\qquad\quad
m_{Ht}^2=\frac{m_Z^2\cos 2\beta\cos(\beta+\alpha)}{\cos(\beta-\alpha)}\,,
\end{equation}
we obtain after considerable simplification,
\begin{equation}
\delta\tan\beta_{{\rm HM}}=\delta\tan\beta_{\overline{\rm DEC}}=\frac{g^2m_D^2}{32\pi^2c_W^2m_Z^2\sin2\beta}\bigg({\frac{1}{\epsilon}-\gamma+\log{4\pi}-\log{\frac{m_M^2}{Q^2}}+1}\bigg)\;.\label{dtbhm}
\end{equation}
Note that the $\mathcal{O}(m_D^2)$ contributions
to the counterterm $\delta\tan\beta$
in the HM and DEC schemes are equivalent, in light of the
absence of non-decoupling terms
in Eq.~(\ref{eq:deltatb}).  Indeed, the $\mathcal{O}(m_D^2)$ contribution
to $\delta\tan\beta$ is independent
of the tree-level Higgs mixing angle $\alpha$.  Although this result
is obvious in the DEC scheme (which is defined via Higgs  wave function
counterterms that are evaluated at $\alpha=0$), the cancellation
of the $\alpha$-dependence in the $\mathcal{O}(m_D^2)$ contribution
to $\delta\tan\beta_{\rm{HM}}$ [defined in~Eq.(\ref{deltatbHM})] is highly non-trivial.

In contrast, in the $\overline{\rm DR}$ scheme only the
$\epsilon^{-1}-\gamma+\log 4\pi$ is retained, so that the
corresponding $\mathcal{O}(m_D^2)$ contribution is simply
\begin{align}
\delta\tan\beta_{\overline{\rm DR}}&=\frac{g^2m_D^2}{32\pi^2c_W^2m_Z^2\sin2\beta}\bigg(\frac{1}{\epsilon}-\gamma+\log{4\pi}\bigg)\,.\label{dtbdr}
\end{align}

\end{document}